\documentclass[8pt]{article}
\usepackage{amsmath,amssymb,amsfonts,amsthm,euscript,changebar,multicol,subfigure,epsfig,amscd,stmaryrd}
\usepackage{graphicx,fancyhdr,color}

\begin{document}



\makeatletter
\newenvironment{figurehere}
  {\def\@captype{figure}}
  {}
\makeatother


\begin{flushleft}
{\huge \fontfamily{phv}\selectfont A resolution of the transition to turbulence paradox} \\[-0.15cm]
\Large \textcolor[rgb]{0.949,0.392,0.003}{Understanding the
transition to turbulence has been impeded by a modeling oversight}
\\[-0.1cm]
\large \textbf{Rouslan Krechetnikov} and \textbf{Jerrold E.
Marsden} (Caltech, Pasadena, CA)
\end{flushleft}

\begin{multicols}{2}

\textbf{Despite being around for over a century, the transition to
turbulence problem remains central in fluid dynamics. This
phenomenon was apparently known to Leonardo da Vinci
\cite{Richter}, who in 1507 introduced the term ``la turbulenza'',
and nowadays it has an impact on practically every field ranging
from astrophysics and atmosphere dynamics to nuclear reactors and
oil pipelines. Beginning with the systematic experimental studies
in a pipe by Osborne Reynolds \cite{Reynolds} in the 1880s, it is
known that the flow becomes turbulent at finite flow rate, usually
measured by Reynolds number $Re=L \, U / \nu$ (see the definitions
in figure \ref{couette}). Similar observations have been made in
other flows, in particular Couette flow -- the flow between two
plates moving parallel to each other, cf. figure
\ref{couette_infinite}, where the transition is observed at
$\mathrm{Re} \simeq 350$ (see, for instance, \cite{Lundbladh}).
Reconciliation of these experimental observations with theory
\cite{Drazin,Romanov} failed because the eigenvalue analysis of
the linearized Navier-Stokes equations (NSEs), which govern fluid
motion, yields eigenvalues $\lambda$ in the left half-plane at all
values of $\mathrm{Re}$, which implies that all small initial
disturbances should decay exponentially like $ e^{\lambda t}$, as
time increases and thus one should have stability. The basic
mathematical setup in classical works \cite{Drazin} treats the
stability problem in an infinite channel $x \in
(-\infty,+\infty)$, as in figure \ref{couette_infinite}. In this
work we demonstrate that this infinite channel assumption is a
sticking point that has prevented one from understanding the
primary instability in the transition to turbulence. Our analysis
on semi-infinite channel-domain $x \in [0,+\infty)$, cf. figure
\ref{couette_semi}, which is more relevant to the way experiments
are usually done, predicts {\it instability} and thus explains
many important features of these phenomena in a simple and basic
way.}

First, let us recall the NSEs, which can be written   as an
evolution equation for the velocity field $\mathbf{u}$,
\begin{align}
\label{main_model} {\mathrm{d} \mathbf{u} \over \mathrm{d} t} = A
\mathbf{u} + \mathbf{N}(\mathbf{u}),
\end{align}
with the linear and nonlinear terms given by
\begin{subequations}
\begin{align}
\label{operator_linearized_NSEs} A \mathbf{u} &= \Bbb{P} \left[-
\mathbf{U} \cdot \nabla \mathbf{u} - \mathbf{u} \cdot \nabla
\mathbf{U} + Re^{-1} \Delta \mathbf{u}\right], \\
\label{operator_nonlinear_NSEs} N(\mathbf{u}) &= - \Bbb{P}
\left[\nabla \cdot (\mathbf{u} \otimes \mathbf{u})\right],
\end{align}
\end{subequations}
where $\Bbb{P}$ is the projection on the space of divergence free
vector fields and $\mathbf{U}$ is the base flow. In the class of
weak solutions, Romanov \cite{Romanov} established linear and
nonlinear stability, based on the absence of linearly unstable
eigenmodes, of the Couette base flow $\mathbf{U} = (y,0)$ on an
infinite domain $x \in (-\infty,+\infty)$. This, of course, leads
to a conceptual difficulty since instability is observed in
experiments.

\begin{figurehere} \centering
\subfigure[Infinite channel setup, $x \in
(-\infty,+\infty)$.]{\epsfig{figure=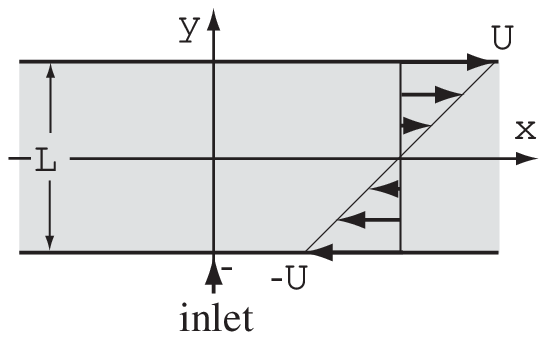,height=1.25in}\label{couette_infinite}}
\subfigure[Semi-infinite channel setup, $x \in
[0,+\infty)$.]{\epsfig{figure=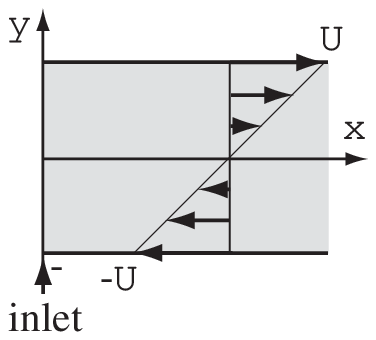,height=1.25in}\label{couette_semi}}
\caption{Couette flow of a fluid with kinematic viscosity $\nu$ in
a channel of width $L$; the base flow is $\mathbf{U} = (y,0)$.}
\label{couette}
\end{figurehere}

The failure of the hydrodynamic stability theory based on
\eqref{main_model} to predict the transition to turbulence
motivated various alternative explanations, including the idea of
a very small basin of attraction of the stable base flow and the
transient growth idea \cite{Trefethen:I,Henningson:I}. It is also
understood that the transition to turbulence belongs to a general
class of counter-intuitive dissipation-induced instabilities based
on the recent theory \cite{Lin,Krechetnikov:I,Krechetnikov:II}.
While all these approaches are still in development, it is worth
mentioning the line of logic of the transient growth concept.
Namely, based on the ansatz that the nonlinear terms
\eqref{operator_nonlinear_NSEs} of the NSEs \eqref{main_model} are
energy conserving and since the linear terms
\eqref{operator_linearized_NSEs} can produce energy only
transiently in time, then the transient growth is the only
explanation of the fact that we observe non-zero deviations from
the laminar base state $\mathbf{U}$ in the aforementioned flows.
The transient growth itself is related to the sensitive and
non-normal nature of \eqref{operator_linearized_NSEs}
\cite{Trefethen:I}, i.e. $A \, A^{*} \neq A^{*} \, A$, where
$A^{*}$ is the operator adjoint to $A$. While this picture has
been appended with various dynamical systems scenarios
\cite{Eckhardt}, such as chaotic saddles, and it also appears to
be useful as a transient effect \cite{Bamieh}, there is still no
theory which would be able to predict a transition robustly. Here
we propose a more direct resolution of this long-standing problem
by demonstrating the existence of linearly unstable eigenmodes.

Before giving a resolution of the mismatch of theory and
experiment, we would like to remind the reader that the fact of
instability implies an existence of at least one eigenmode
$f_{\lambda}(x)$ such that the corresponding eigenvalue $\lambda$
has a positive real part. Perhaps because of the translational
invariance of the base state $\mathbf{U}=(y,0)$, cf. figure
\ref{couette_infinite}, or for convenience, the stability of
Couette flow has always been studied on an infinite domain,
$-\infty < x < +\infty$. However, if one recalls the way the
experiments on the transition are usually done, i.e. one
introduces disturbances at some fixed inlet location, say $x=0$,
and observes how they evolve downstream, then it becomes clear
that the semi-infinite domain, $x \in [0,+\infty)$, as in figure
\ref{couette_semi} is more relevant as a mathematical
idealization. Then, it is convenient to study the linear
eigenvalue problem -- the classical Orr-Sommerfeld (OS) equation
-- by assuming that the disturbance eigenfunction is of the form
$\sim a_{\lambda \mu}(y) e^{\lambda t} e^{- \mu x}$, where
$\lambda \in \Bbb{C}$ is the eigenvalue and $\mu \ge 0$ is an
analog of the wavenumber:
\begin{subequations}
\label{Orr_Sommerfeld_Couette}
\begin{align}
\label{Orr_Sommerfeld_Couette_equation} \left[{\mathrm{d}^{2}
\over \mathrm{d}y^{2}} + \mu^{2} - Re \left(\lambda - \mu y\right)
\right] \left({\mathrm{d}^{2} \over \mathrm{d}y^{2}} +
\mu^{2}\right) a(y) = 0, \\
\label{Orr_Sommerfeld_Couette_BCs} y=-1,1: \, a = a_{y} = 0,
\end{align}
\end{subequations}
where we dropped the indeces $\lambda$ and $\mu$. The basis
functions $e^{- \mu x}$, $\mu \ge 0$, are clearly not members of
the space of bounded functions on the whole real line, $x \in
(-\infty,+\infty)$, which are used in the classical analysis of
this problem, but they do belong to the space of functions bounded
on the semi-infinite domain, $x \in [0,+\infty)$. Therefore, these
eigenfunctions were not captured in the traditional approach.
Alternatively, equation \eqref{Orr_Sommerfeld_Couette} could be
treated with cosine and sine Fourier-transforms, which would lead
to the same results but with a considerably more complicated
version of \eqref{Orr_Sommerfeld_Couette}. We choose to work with
exponential functions $e^{- \mu x}$, since our goal here is to
demonstrate the existence of unstable eigenmodes in a direct way;
other reasons will be clear from the subsequent discussion. Also,
this choice of functions leads to the OS equation analogous to the
classical one \cite{Drazin}: the only difference of
\eqref{Orr_Sommerfeld_Couette} from the classical case of the OS
equation \cite{Drazin} studied on an infinite domain is $\mu$
versus $-i k$, where $k$ is the wavenumber.

\begin{figurehere}
\centering \subfigure[$\mathrm{max} \, Re(\lambda)$; $\circ$:
$\mathrm{Re}=10^{4}$, $\times$: $\mathrm{Re}=10^{3}$, $\square$:
$\mathrm{Re}=10^{2}$, $+$:
$\mathrm{Re}=10^{1}$.]{\epsfig{figure=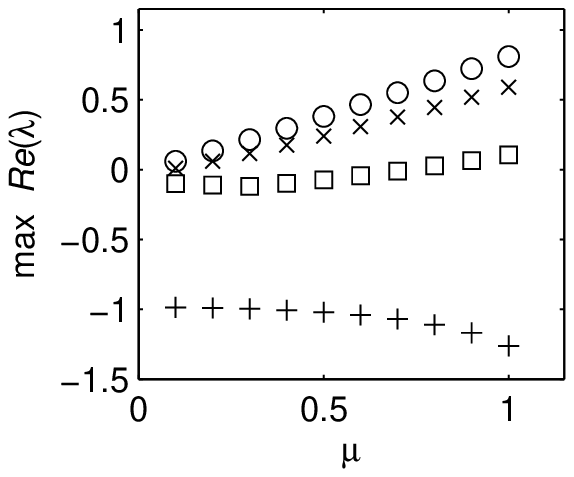,height=1.75in}\label{lambda_mu}}
\subfigure[Critical Reynolds number
$\mathrm{Re}_{c}(\mu)$.]{\epsfig{figure=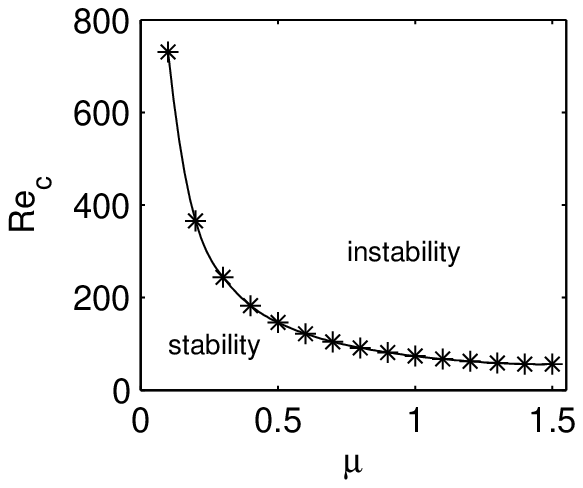,height=1.79in}\label{Re_c_mu}}
\subfigure[Asymptotics $\mathrm{max} \, Re(\lambda)$ for
$\mathrm{Re}=10^{4}$.]{\epsfig{figure=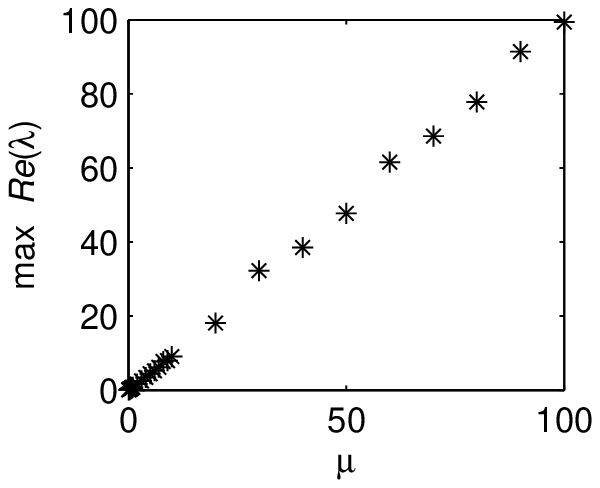,height=1.75in}\label{lambda_asymptotic}}
\subfigure[Spectrum: distribution of eigenvalues $\lambda_{n}$, $n
\in \Bbb{N}$ of \eqref{Orr_Sommerfeld_Couette} in the complex
plane (eigenvalues continue to the negative part of the real axis)
for $\mu=1$ and $Re=70$ (red dots), $Re=1000$ (green dots), and
$Re=5000$ (blue
dots).]{\epsfig{figure=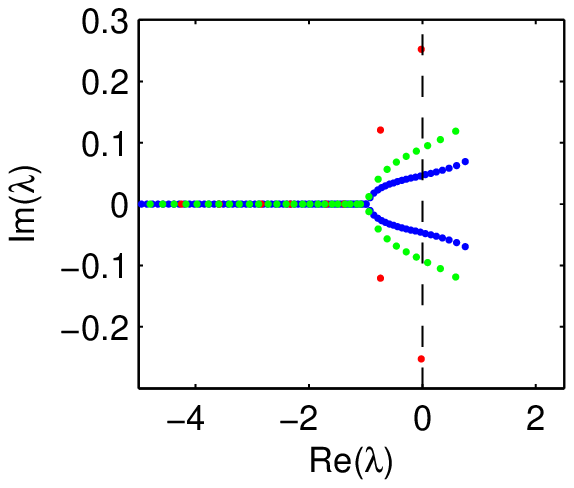,height=1.75in}\label{spectrum_Re5000mu1}}
\caption{Stability picture of the Couette flow.}
\label{stability_picture}
\end{figurehere}

The eigenvalue problem \eqref{Orr_Sommerfeld_Couette} is solved
numerically by expanding $a(y)$ in Chebyshev polynomials
\cite{Orszag}. As one learns from figure \ref{stability_picture},
the stability picture on a semi-infinite domain is in sharp
contrast to what one has on an infinite domain \cite{Drazin}, but
conforms well with the usual intuitive understanding of
instability phenomena: for some values of $\mu$ there are
eigenvalues with positive real part and thus Couette flow is
absolutely unstable. In fact, since in general all values of $\mu$
may be present in a real flow, then figure \ref{lambda_mu}
suggests that the transition in the Couette flow is not a critical
phenomenon. Indeed, figure \ref{lambda_asymptotic} indicates that
the instability in the Couette flow is in fact a short-wave
instability, since the value of $\mathrm{max} \, Re(\lambda)$ is
increasing with $\mu$, and larger $\mu$ means that the disturbance
is localized around the inlet.  It is also notable that the
structure of the eigenfunctions corresponding to the leading
eigenvalues increases in complexity with increasing $\mu$, as
illustrated in figure \ref{eigenfunctions}, which may explain the
tangled flow picture observed experimentally in the Couette flow:
see \cite{Andersson} and references therein. However, if in a
particular experiment, the admissible magnitudes of $\mu$ are
restricted to a range of small values, then one can observe
critical phenomena, as in figure \ref{Re_c_mu}. These are clearly
of Hopf bifurcation type, common in various fluid dynamics
problems \cite{Marsden}, as follows from the distribution of
eigenvalues in the complex plane in figure
\ref{spectrum_Re5000mu1}, which illustrates that the leading
(rightmost) eigenvalues cross the imaginary axis as Reynolds
number increases. In this case one can expect that the leading
eigensolutions are the usual Tollmien-Schlichting waves
\cite{Lin,Drazin} appearing via the Hopf bifurcation. We have to
stress, however, that in general, if all values of $\mu$ are
present, then the transition to turbulence is not a critical
phenomenon and thus not a Hopf bifurcation, similar to the
Rayleigh-Taylor instability, i.e. instability of a heavy fluid
accelerating into a light one.

\begin{figurehere}
\centering \epsfig{figure=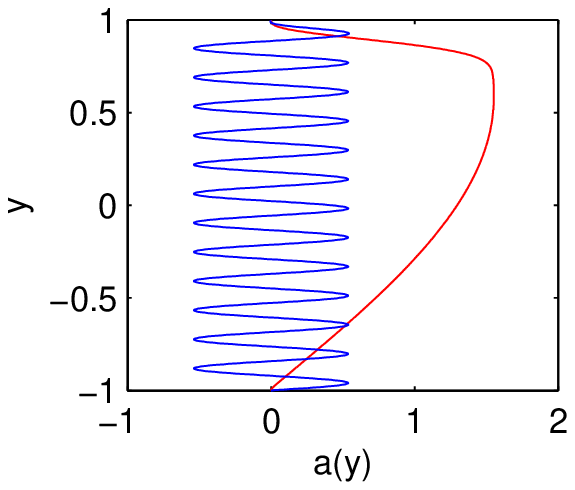,height=1.75in}
\caption{The eigenfunctions corresponding to the rightmost
eigenvalue for the Couette flow at $Re=5000$: red curve
corresponds to $\mu=1$ and $\mathrm{Re} \, \lambda_{\mathrm{max}}
\simeq 38.53$, blue curve corresponds to $\mu=40$ and $\mathrm{Re}
\, \lambda_{\mathrm{max}} \simeq 0.76$.} \label{eigenfunctions}
\end{figurehere}

There is much more to this stability picture and a lot remains to
be understood about the properties of equation
\eqref{Orr_Sommerfeld_Couette}, as well as a full
function-analytic nonlinear analysis of \eqref{main_model} with
careful treatment of the inlet boundary conditions is needed
similar to \cite{Krechetnikov:II}. However, the key feature -- the
existence of unstable eigenmodes -- originating from the
semi-infinite domain setup is well illustrated above. Analogous
computations performed for the plane Poiseuille flow also revealed
a similar instability picture, which suggests that the right
mathematical setup used above is a universal explanation of the
transition to turbulence in the aforementioned ``troublesome''
flows.

This counter-intuitive difference in the stability results between
the semi-infinite and infinite domains can be appreciated with the
assistance of the sketch in figure \ref{semi_vs_infinite}.

\begin{figurehere} \centering
\epsfig{figure=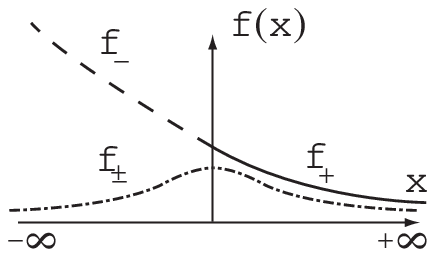,height=1.25in}
\caption{Eigenmodes in the stability analysis: semi- versus
infinite domain.} \label{semi_vs_infinite}
\end{figurehere}

Namely, if, for example, one restricts (eigen-) functions $f(x)$
to be bounded for all $x$ as dictated by the fact that the
physical solution should be bounded, then eigenfunctions defined
on $x \in (-\infty,+\infty)$ are more restrictive compared to
eigenfunctions defined on $x \in [0,+\infty)$. Indeed, if one can
construct a function $f_{+}$ bounded on $x \in [0,+\infty)$ which
also satisfies the OS equation
\eqref{Orr_Sommerfeld_Couette_equation}, then continuation of this
function onto $x \in (-\infty,0]$ may lead to an unbounded
function $f_{-}$, as dictated by the structure of the linear
operator \eqref{operator_linearized_NSEs} and as illustrated in
figure \ref{semi_vs_infinite}. Our exponential eigenfunctions
$e^{-\mu \, x}$ are a good example of functions bounded on the
right half-line and unbounded on the left half-line, while still
satisfying the OS equation \eqref{Orr_Sommerfeld_Couette}. These
observations can be illustrated with the following elementary
eigenvalue problem:
\begin{align}
\lambda {\mathrm{d}^{2} \phi(x) \over \mathrm{d} {x}^{2}} +
{\mathrm{d}^{3} \phi(x) \over \mathrm{d} {x}^{3}}  = 0,
\end{align}
which clearly contrasts the problems on infinite and semi-infinite
domains:
\begin{itemize}
\item $x \in (-\infty,+\infty)$ and
$\phi(x) \in L_{2}$: applying Fourier transform we get $\lambda =
- i k$, $k \in \Bbb{R}$, i.e. marginal stability.
\item $x \in [0,+\infty)$ and $|\phi(x)| < \infty$: instability
is present since there are eigenfunctions $\phi \sim e^{- \mu x}$,
$\mu \ge 0$, with eigenvalues $\lambda = \mu$.
\end{itemize}

Note that the above arguments also explain the sensitivity of the
experimentally observed critical Reynolds number $Re_{c}$ to the
properties of disturbances at the domain inlet, $x=0$: while their
amplitudes do not play a role in view of the linearity of the
problem, gradient-like properties of the disturbances do! Namely,
by varying gradient-like properties of disturbances at $x=0$
effectively changes the boundary conditions at $x=0$ and thus the
size of the eigenspace. Since restricting the domain to $x \in
[0,+\infty)$ enlarges the function space, one can expect that the
spectrum enlarges as well and may lead to instabilities. This
simple fact, as we saw above, explains the mechanism behind the
transition to turbulence! While the above analysis has been
conducted on a semi-infinite domain with open inlet and outlet
boundaries, which demonstrated the existence of absolutely
unstable eigenmodes, the latter apparently also exist on finite
length domains with open inlet and outlet boundaries.

Finally, note that there is nothing unusual about increasing
growth rate for increasing $\mu$, observed in figure
\ref{lambda_asymptotic}, which is a common feature in many
fundamental hydrodynamic instabilities in the short-wave limit,
such as the Rayleigh-Taylor instability. Of course, in the
nonlinear setting one does not observe infinite growth rates,
because they are suppressed by the nonlinear terms, which play a
stabilizing role opposing the formation of singularities even in
the case when they are energy conserving \cite{Hou}; in our case
the nonlinear terms can also be dissipative. The latter fact is in
apparent contrast with the above mentioned ansatz of the transient
growth story, i.e. that the nonlinear terms are energy-conserving.
This ansatz is valid if the disturbance field in the Couette flow
problem is considered on an infinite domain with the boundary
conditions at $x = \pm \infty$ corresponding to decay to zero, but
not on a semi-infinite domain. Indeed, multiplying
\eqref{main_model} with $\mathbf{u}^{T}$ and integrating over the
semi-infinite strip-like flow domain $\Omega$ we arrive at the
Reynold-Orr equation for the kinetic energy
$\label{kinetic_energy} E(t) = \|\mathbf{u}\|^{2} / 2$:
\begin{align}
\label{equation_RO_full} &- {\mathrm{d} E \over \mathrm{d} t} =
\int\limits_{\partial \Omega}{n_{i} u_{i} p \, \mathrm{d}s} + \nu
\int\limits_{\partial \Omega}{n_{j} u_{i}
u_{i,j} \, \mathrm{d}s} + \nu \|\nabla \mathbf{u}\|^{2} \\
&+ \left<D_{ij},u_{i} u_{j}\right> + \boxed{{1 \over 2}
\int_{\partial \Omega}{n_{j} u_{j} u_{i} u_{i} \, \mathrm{d}s}} +
{1 \over 2} \int_{\Omega}{n_{j} U_{j} u_{i} u_{i} \, \mathrm{d}x},
\nonumber
\end{align}
where $\mathbf{n}$ is the outward normal of $\partial \Omega$. If
the domain $\Omega$ is unbounded and open, as in the Couette or
pipe flows, then the effect of the nonlinear terms (cubic term in
\eqref{equation_RO_full}) does not disappear, since disturbances
are non-zero at the inlet and, if they lead to an instability, do
not necessarily decay at infinity.

\bibliographystyle{unsrt}

\end{multicols}

\end{document}